\documentclass[pra,aps,twocolumn,english]{revtex4}
\usepackage{graphicx}
\usepackage{amssymb, amsmath, color, rotating}

\begin{document}

\title{Spin waves in a spin-$1$ normal Bose gas}

\author{Stefan S. Natu}
\email{ssn8@cornell.edu}
\author{Erich J. Mueller}
\email{em256@cornell.edu}
\affiliation{Laboratory of Atomic and Solid State Physics, Cornell University, Ithaca, New York 14853, USA.}

\begin{abstract}
We present a theory of spin waves in a non-condensed gas of spin-1 bosons: providing both analytic calculations of the linear theory, and full numerical simulations of the nonlinear response.  We highlight the role of spin-dependent contact interactions in the dynamics of a thermal gas. Although these interactions are small compared to the thermal energy, they set the scale for low energy long wavelength spin waves. In particular, we find that the polar state of $^{87}$Rb is unstable to collisional mixing of  magnetic sublevels even in the normal state. We augment our analytic calculations by providing full numerical simulations of a trapped gas, explicitly demonstrating this instability. Further we show that for strong enough anti-ferromagnetic interactions, the polar gas is unstable. Finally we explore coherent population dynamics in a collisionless transversely polarized gas.  \end{abstract}
\maketitle

\section{Introduction}To date, all spin wave experiments in non-condensed systems have been on spin-$\frac{1}{2}$ or pseudo-spin-$\frac{1}{2}$ systems. Here we ask the following questions:  what is the nature of spin waves in a thermal spin-$1$ gas? How do these differ from the well understood spin-$\frac{1}{2}$ case? These questions are of experimental interest, with several groups possessing the technology to study them using ultra-cold gases \cite{mukund, chapman, sengstock}. 

In the context of cold gases, spin waves were first discussed by Bashkin \cite{bashkin}, and independently by Lhuillier and Lalo\"e \cite{laloe}. The key finding was that spin exchange collisions can give rise to weakly damped spin waves even in a non-degenerate gas. Ultra cold gases have provided an exciting setting for observing these spin phenomena. In particular, experiments on pseudo spin-$\frac{1}{2}$ Bose and Fermi systems by the JILA \cite{cornell} and Duke \cite{du} groups have observed coherent collective oscillations in an otherwise classical gas.

One expects the physics of a spin-$1$ gas to be far richer than a pseudo-spin-$\frac{1}{2}$ system. This is amply demonstrated by experiments on condensed spin-$1$ gases \cite{mukund2, sengstock2, chapman3, chapman4}. One of the most dramatic observations was that of a dynamical instability in Bose condensed $^{87}$Rb, studied by Sadler \textit{et al.} \cite{sadler}. Beginning with a gas pumped into the $m_{F} = 0$ state, they observed the spontaneous formation of transverse ferromagnetic domains. Will a similar instability be observed in the normal state? We find that there is an exponentially growing mode in the unmagnetized gas, even in the normal state, with a wavelength comparable to typical cloud sizes. 

From a theoretical perspective, the source of novel physics in a dilute spin-$1$ Bose gas (such as $^{87}$Rb, $^{23}$Na) is the structure of the interactions, described by \textit{two} coupling constants: $c_{0}$ and $c_{2}$, representing spin-independent and spin-dependent contact interactions.  The interaction Hamiltonian density takes the form $H_{\rm int} = c_0 n^2/2 + c_2 \langle \vec{\textbf{S}} \rangle^2/2$, where $n$ and $\vec{\bf S}$ are the local density and spin density respectively \cite{jason}.  The coefficient $c_{2}$ has no corresponding analog in the spin-$\frac{1}{2}$ case. This interaction gives rise to spin mixing collisions, where two atoms in the $m_{F} = 0$ hyperfine sub-level can combine to form atoms in the $m_{F} = \pm 1$ states \cite{ketterle}. Another important consideration is the quadratic Zeeman effect, which arises from the hyperfine interaction and the difference in the coupling between the electronic and nuclear spins.  The linear Zeeman effect can be neglected in the Hamiltonian as the total spin of the atoms is a conserved quantity. 

We begin our analysis by setting up the problem, and reviewing spin waves in a spin-$\frac{1}{2}$ gas. This allows us to highlight the differences with the spin-$1$ case. Next, we turn to the spin-$1$ gas. Starting from a microscopic Hamiltonian, we obtain a linearized Boltzmann equation about the ferromagnetic ($m_{F} = 1$) and polar ($m_{F} = 0$) states, and calculate the spin wave dispersion in each case. We find that a polar gas with ferromagnetic interactions is dynamically unstable towards spin-mixing collisions for small enough Zeeman fields, analogous to the condensed case. By explicit calculation of the dispersion relation, we show that for strong enough anti-ferromagnetic interactions, an instability occurs in the polar state \cite{mueller2}. 

Following these analytic calculations, we perform numerical simulations of a trapped gas using an effective $1$D Boltzmann equation. We explore the evolution of transversely polarized spins, and investigate dynamical instabilities.

Our work complements prior work on the kinetics of a normal spin-$1$ Bose gas by Endo and Nikuni \cite{nikuni}. While their focus is on the effect of collisions in the damping of collective modes of a trapped gas, with particular emphasis on dipole modes, we focus here on the collisionless, Knudsen regime. 

\section{Basic Setup}
\subsection{Kinetic Equations} As described by Ho \cite{jason} and Ohmi and Machida \cite{Ohmi}, the second-quantized Hamiltonian for a spin-$1$ gas, expressed in a frame where each spin component is rotating at its Larmor frequency is
\begin{widetext}
\begin{eqnarray}\label{eq:1}
{\cal H} = \int d\vec{\textbf{r}}~ \psi_{a}^{\dagger}\left(-\frac{1}{2m} + U( \vec{r}, t) + q \textbf{S}_{z} \cdotp \textbf{S}_{z}\right)\psi_{a} + \frac{c_{0}}{2}\psi^{\dagger}_{a}\psi_{a^{'}}^{\dagger}\psi_{a^{'}}\psi_{a} 
+ \frac{c_{2}}{2}\psi^{\dagger}_{a}\psi^{\dagger}_{a^{'}}\vec{\textbf{S}}_{ab}\cdotp \vec{\textbf{S}}_{a^{'}b^{'}}\psi_{b^{'}}\psi_{b}
\end{eqnarray}
\end{widetext}
where $a = (-1, 0, 1)$, is the quantum number for the z-component of the spin, $\psi_{\sigma}(\textbf{r})$ is the field annihilation operator obeying bosonic commutation relations, and $U (\vec{r}, t)$ is the trapping potential. Here $\vec{\textbf{S}}$ denotes the dimensionless vector spin operator. Throughout, we set $\hbar = 1$. Boldface is used to denote matrices, and arrows denote vectors.

The interaction strengths, expressed in terms of the scattering lengths in the spin-$2$ and $0$ channel ($a_{2}, a_{0}$) are $c_{0} = 4\pi(a_{0} + 2a_{2})/3m$ and $c_{2} = 4\pi(a_{2} - a_{0})/3m$. A negative $c_{2}$  favors a ferromagnetic state with $\langle \vec{\textbf{S}} \rangle = 1$, while for positive $c_{2}$, the equilibrium state is one with $\langle \vec{\textbf{S}} \rangle = 0$, where all the atoms are in the $m_{F} = 0$ state, or simply an incoherent mixture.

Additionally, one considers the quadratic Zeeman effect $q \propto B^{2} $, which favors a state with $m_{F} = 0$ ($\langle \vec{\textbf{S}}\cdotp \vec{\textbf{S}} \rangle = 0$). Thus for a gas with negative $c_{2}$ such as $^{87}$Rb, the spin dependent contact interaction competes with the quadratic Zeeman field, giving rise to interesting dynamics \cite{sadler}. 

In the condensed gas, dipolar interactions may also be important \cite{mukund, mukund2}. At the lower densities found in a normal gas these interactions, which fall as $1/r^{3}$, may be neglected.

Following standard arguments \cite{kadanoff} we obtain the equations of motion for the Wigner density matrix $\textbf{F}_{ab}(\vec{p}, \vec{R}, t)$, whose elements are $f_{ab}(\vec{p}, \vec{R}, t)= \int d \vec{r} e^{-i\vec{p}\cdotp \vec{r}} \langle \psi^{\dagger}_{a}(\vec{R} - \frac{\vec{r}}{2}, t) \psi_{b}(\vec{R} + \frac{\vec{r}}{2}, t) \rangle$. The diagonal components of the spin density matrix, when integrated in momentum, give the densities of each of the spin species. The off diagonal terms, often referred to as \textit{coherences}, are responsible for spin dynamics. Here $\vec{p}$ represents momentum, $\vec{R}$ and $\vec{r}$ denote the center of mass and relative coordinates. The Wigner function is the quantum analog of the classical distribution function. By taking moments of the Wigner function, we obtain physical observables such as the density $\textbf{n}(\vec{R}, t) = \int \frac{d\vec{p}}{(2 \pi)^{3}}~\textbf{F}(\vec{p}, \vec{R}, t)$ and spin current $\vec{\textbf{j}}(\vec{R}, t) = \int \frac{d\vec{p}}{(2 \pi)^{3}}~\vec{p}~\textbf{F}(\vec{p}, \vec{R}, t)$.

The equation of motion takes the standard form \cite{bashkin}:
\begin{equation}\label{eq:2}
\frac{\partial}{\partial t}\textbf{F} + \frac{\vec{p}}{m}\cdotp\vec{\nabla}_{R}\textbf{F}= i\Big[\textbf{V}, \textbf{F}\Big] + \frac{1}{2}\Big\{\vec{\nabla}_{R}\textbf{V},\vec{\nabla}_{p}\textbf{F}\Big\} + \textbf{I}_{c}
\end{equation}
where $\textbf I_{c}$ is the collision integral and $\textbf V$ is the interaction potential. This form of the Boltzmann equation (\ref{eq:2}) is completely general, and holds for any non-condensed spinor gas. The role of spin enters in determining the dimension of the density matrix, and the exact form of the interaction potential $\textbf{V}$. 

The first two terms on the right hand side of the Boltzmann equation (\ref{eq:2}) arise from forward and backward scattering collisions. While the former type merely alter the mean field seen by the atoms, backward scattering collisions allow the colliding atoms to exchange momentum. 

The last term in the Boltzmann equation is the collision integral, responsible for energy relaxation. While a detailed derivation of the collision integral is non trivial \cite{nikuni}, the qualitative properties are well described within a simple relaxation time approximation $I_{c} = -(f - f_{0})/\tau$. The relaxation time ($\tau$) is proportional to the elastic scattering rate $\tau^{-1}_{el} \sim 8\pi a^{2}_{0}v_{T}n$, where $v_{T} = \sqrt{\frac{2 k_{B}T}{m}}$ is the thermal velocity and $n$ is the density. 

The precise expression for $\bf{V}$ determined by including the interactions within the Hartree-Fock description is\begin{equation}\label{eq:3}
 \textbf{V} = \left(U + c_{0}\text{Tr}(\textbf n)\right)\textbf 1 + c_{0}\textbf n + c_{2}\vec{\textbf{S}}\textbf n \cdotp \vec{\textbf{S}} + c_{2} \vec{M}\cdotp\vec{\textbf{S}} 
\end{equation}
where $\textbf{1}$ is the identity matrix, Tr denotes the trace operation, and $\vec{\textbf{M}} = \text{Tr}(\vec{\textbf{S}}~\textbf{n})$ is the magnetization. Our form for the interaction energy is equivalent to that of Endo and Nikuni \cite{nikuni}. One may explicitly check that this interaction potential is rotationally invariant in spin space.

In experiments the external trapping potential $U$ is often spin-dependent, an effect which is readily incorporated. The spin independent interaction gives rise to self and cross interaction terms. The latter give rise to coherences and are encoded in the second term in (\ref{eq:3}). In addition, the mean-field potential alters the trapping potential seen by all the atoms by an amount $c_{0}n_{tot}$, where the total density $n_{tot} = \text{Tr}(\textbf{n})$.

The contribution of the spin-dependent interaction is more subtle and can be understood as follows: The first term accounts for spin dynamics such as spin-relaxation collisions. For ferromagnetic interactions ($c_{2} < 0$), the last term increases the density of regions where the atomic spins are aligned with respect to one another ($ |M| = 1$). For a fully polarized gas in the $m_{F} = +1$ sublevel, this is $c_{2}\textbf{S}_{z}$, while for an unmagnetized ($m_{F} = 0$) gas it is zero. 

Finally, we note that while the self- and cross- interactions between the three sublevels produce diagonal and off-diagonal contributions to the interaction potential, the spin-relaxation or population exchange interactions only gives rise to coherences, and are absent for a single component gas. 

\subsection{Qualitative Features}

Here we elaborate on our argument for why one expects richer spin physics in a spin-$1$ system versus a spin-$\frac{1}{2}$ one. We assume a uniform, collisionless gas with 
$\nabla_{R}c_{0}n/m v_{T} \ll c_{0}n$, such that all the physics is governed by the commutator in (\ref{eq:2}). 

The pseudo-spin $\frac{1}{2}$ case may be understood starting from the fact that that a $2 \times 2$ matrix ($A$) can be decomposed into $\textbf{A} = A_{0}\textbf{I} + A_{\mu}\sigma_{\mu}$, where $A_{0} = \text{Tr}(\textbf{A})$, and $\vec{\sigma}$ are the Pauli matrices.  Expressing the density matrix and interaction potential in this way, the equations of motion are $\frac{\partial}{\partial t}F_{0} + \frac{\vec{p}}{m}\cdotp\vec{\nabla}_{R}F_{0} = 0$ and $\frac{\partial}{\partial t}{F_{\mu}} + \frac{\vec{p}}{m}\cdotp\vec{\nabla}_{R}F_{\mu} =  - F_{\rho}V_{\nu}\epsilon_{\rho\nu\mu}$, where $\epsilon_{\rho\nu\mu}$ is the completely antisymmetric unit tensor. The second equation, which is responsible for much of the spin wave physics in a spin-$\frac{1}{2}$ system simply says that the interactions act as an effective magnetic field about which the spins precess.

To analyze the spin-$1$ case, we follow Ohmi and Machida \cite{Ohmi} and use a Cartesian basis ($\psi =\{ \psi_{x}, \psi_{y}, \psi_{z}\}$). This representation is related to the spherical ($\{1, 0, -1\}$) basis as follows \cite{mueller}: $\psi_{x} = \frac{1}{\sqrt{2}}(\psi_{1} - \psi_{-1})$,$\psi_{x} = \frac{i}{\sqrt{2}}(\psi_{1} + \psi_{-1})$ and $\psi_{z} = \psi_{0}$. 

In the Cartesian basis the irreducible decomposition of a spin-$1$ system is $\textbf{A} = A_{0}\textbf{I} + i \epsilon_{abc}A^{(1)}_{c} + A^{(s)}$, where the scalar $A_{0} = \frac{\text{Tr}(A)}{3}$, $\textbf{A}^{(a)} = \epsilon_{abc}A^{(1)}_{c}$ is a completely antisymmetric matrix proportional to the vector spin $\langle\textbf{S}\rangle$ order, and $\textbf{A}^{(s)}$ is a symmetric traceless tensor which is related to the spin fluctuations $\langle \textbf{S}_{a}\textbf{S}_{b} + \textbf{S}_{b}\textbf{S}_{a} \rangle$, and is a nematic degree of freedom. 

Writing the density matrix ($\textbf{F}$) and interaction ($\textbf{V}$) in terms of their respective irreducible decompositions we obtain three equations of motion: $\frac{\partial}{\partial t}F_{0} + \frac{\vec{p}}{m}\cdotp\vec{\nabla}_{R}F_{0} = 0$, $\frac{\partial}{\partial t}{F^{(a)}} + \frac{\vec{p}}{m}\cdotp\vec{\nabla}_{R}F^{(a)} = i[\textbf{V}^{(a)}, \textbf{F}^{(a)}] + i[\textbf{V}^{(s)}, \textbf{F}^{(s)}]$, and $\frac{\partial}{\partial t}{F^{(s)}} + \frac{\vec{p}}{m}\cdotp\vec{\nabla}_{R}F^{(s)} = i[\textbf{V}^{(a)}, \textbf{F}^{(s)}] + i[\textbf{V}^{(s)}, \textbf{F}^{(a)}]$. Here the second equation describes the evolution of $\langle \textbf{S} \rangle$ and the spin currents, while the last equation describes dynamics of the nematicity ($\textbf{A}^{(s)}$).

The new feature is that the spin and nematic degrees of freedom are coupled by the interaction matrix. Using the Levi-Civita identities, it is easy to see that $i[\textbf{V}^{(a)}, \textbf{F}^{(a)}] $ is the spin-$1$ analog of the corresponding spin-$\frac{1}{2}$ term which acts as an effective magnetic field for the spins. But the last term $i[\textbf{V}^{(s)}, \textbf{F}^{(s)}]$ shows that fluctuations of the nematicity can change the spin dynamics. 

\subsection{Spin waves in a two-component gas} As a starting point for understanding the spin-$1$ gas, we review the spin-$\frac{1}{2}$ case. As explained in \cite{oktel}, spin dynamics in the cold collision regime are governed by momentum exchange collisions with a characteristic timescale $\tau^{-1}_{ex} \sim \frac{4\pi a_{12}n}{m}$, where $a_{12}$ is the two-body scattering length. When $k_{B}T \ll \frac{1}{m a^{2}_{12}}$, or alternatively, the thermal deBroglie wavelength is large compared to the scattering length $\Lambda_{T} = \sqrt{\frac{2\pi}{m k_{B}T}} \ge a_{12}$, $\tau_{ex} \ll \tau_{el}$, and for times shorter than $\tau_{el}$, particles exchange momentum several times without significantly altering the energy distribution. Therefore excitations with wavelength longer than $v_{T}\tau_{ex}$ are a 
collective effect. 

For typical densities ($\sim 10^{14}$cm$^{-3}$) and scattering lengths,  $n^{1/3}a \sim 0.01$ the condensation temperature ($T_{c}$)  for the interacting system may be approximated to that of an ideal Bose gas \cite{baym}. Thus in order to see collective spin phenomena, we require $T_{c} \sim \langle \omega \rangle N^{\frac{1}{3}} < k_{B}T < \frac{1}{m a^{2}_{12}}$, where $\langle \omega \rangle = (\omega_{x}\omega_{y}\omega_{z})^{1/3}$ is the average trap frequency \cite{pethick}. Although this is typically a wide temperature range $10^{-7} < T < 10^{-3}$K, diffusive relaxation damps out spin waves at higher temperatures.

A two-component gas has a longitudinal and a transverse spin mode. The longitudinal mode has a linear dispersion, but is strongly Landau damped \cite{bashkin}. By contrast, the transverse mode is weakly damped and propagates with dispersion $\omega = \frac{k^{2}v^{2}_{T}\tau_{ex}}{2}\left(1 - i \frac{\tau_{ex}}{\tau_{D}}\right)$, where $\frac{\tau_{ex}}{\tau_{D}} \sim \frac{a_{12}}{\Lambda_{T}} \le 1$ \cite{levy}. These weakly damped spin waves have been observed in the context of spin polarized Hydrogen by Johnson \textit{et al.} \cite{cornell, du, lee}. 

\section{Spin waves in the spin-$1$ gas}
In this section we linearize Eq.(\ref{eq:2}) about stationary states. In Sec.IV we consider more general dynamics.

\subsection{Excitations about the $m_{F}=1$ state}We begin by linearizing about the $m_{F} = 1$ state. We start with a homogenous gas of particles with a Maxwellian velocity distribution and initial density $n_{0}$.  The self-interaction between two particles is proportional to $a_{2}$, and so we define $\Omega_{int} = (c_{0} + c_{2})n_{0}$. We begin by considering the collisionless limit,  $1 \ll \Omega_{int}\tau_{D}$.  

\begin{figure}[hbtp]
\begin{picture}(150, 160)(10, 10)
\put(-38, 160){(a)}
\put(92, 160){(b)}
\put(-38, 87){(c)}
\put(92, 83){(d)}
\put(-25, 95){\includegraphics[scale=0.375]{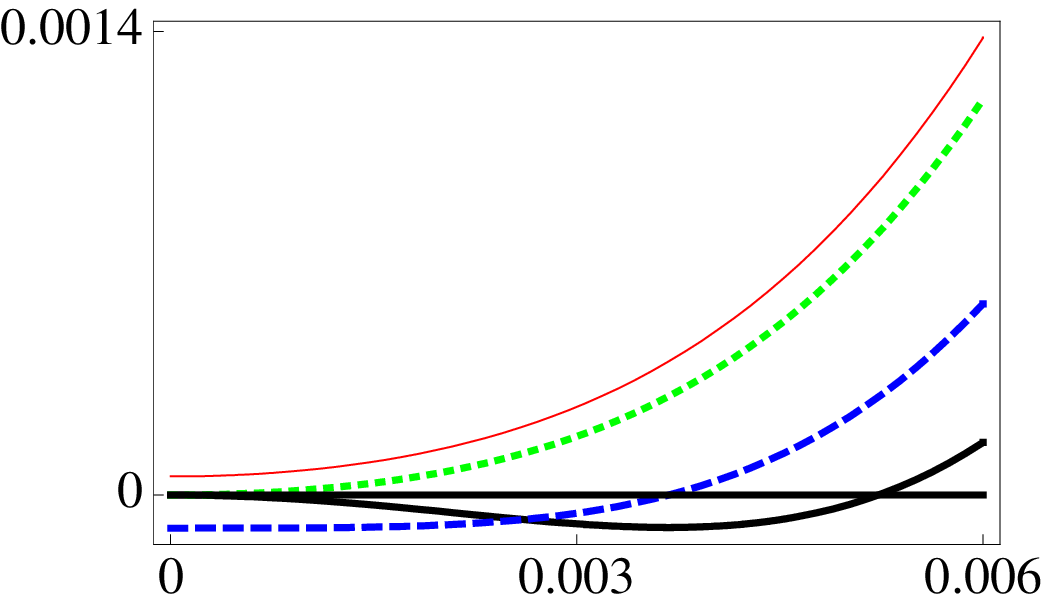}}
\put(95, 95){\includegraphics[scale=0.375]{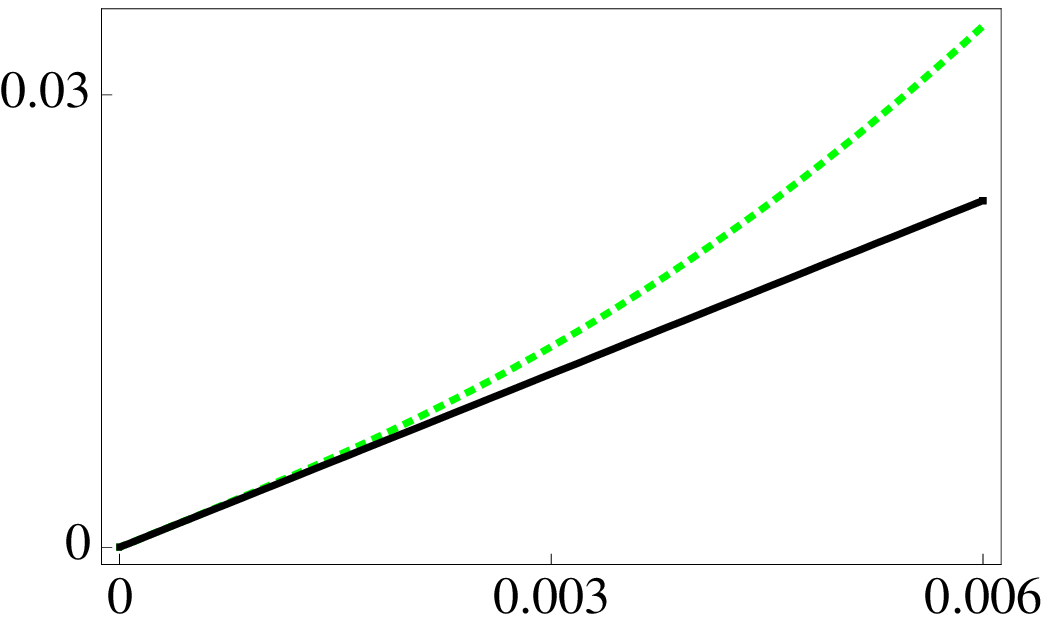}}
\put(-25, 15){\includegraphics[scale=0.375]{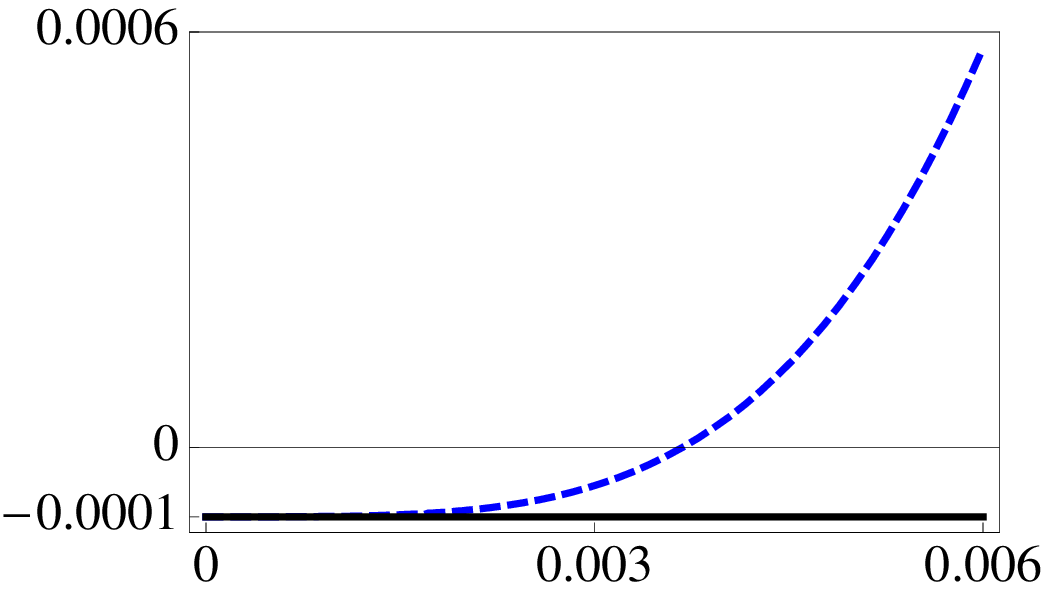}}
\put(95, 15){\includegraphics[scale=0.375]{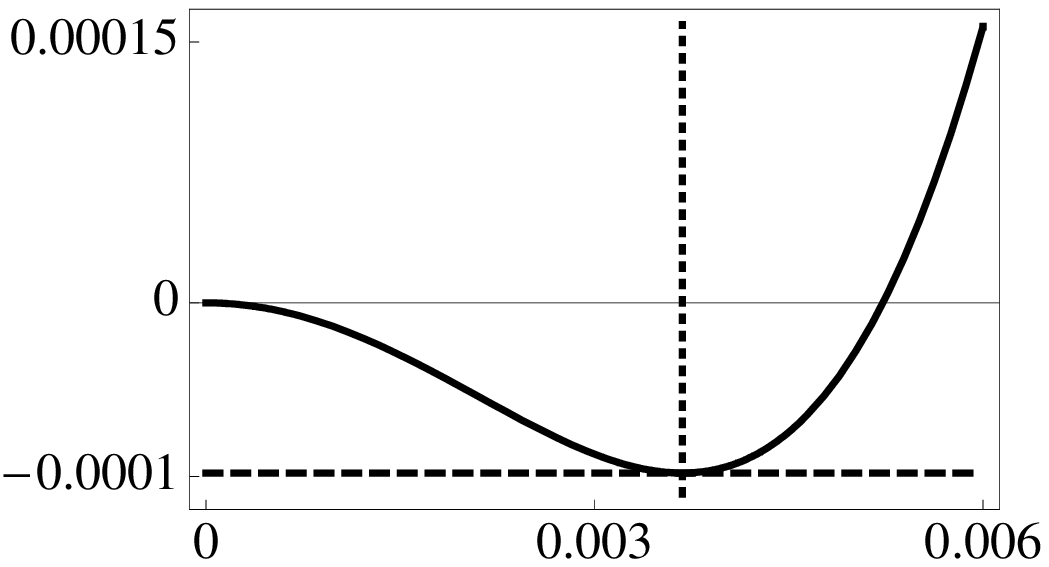}}
\put(-28, 110){\begin{sideways}\textbf{$(\omega/c_{0}n_{0})^{2}$}\end{sideways}}
\put(95, 115){\begin{sideways}\textbf{$\omega/c_{0}n_{0}$}\end{sideways}}
\put(-30, 35){\begin{sideways}\textbf{$(\omega/c_{0}n_{0})^{2}$}\end{sideways}}
\put(92, 30){\begin{sideways}\textbf{$(\omega/c_{0}n_{0})^{2}$}\end{sideways}}
\put(155, 87){\textbf{$k \Lambda_{T}$}}
\put(35, 87){\textbf{$k \Lambda_{T}$}}
\put(155, 7){\textbf{$k \Lambda_{T}$}}
\put(35, 7){\textbf{$k \Lambda_{T}$}}
\end{picture}
\caption{\label{fig:-1}(Color Online) (a) Dispersion relation $\omega^2(k)$ for various values of $q$ using $^{87}$Rb parameters ($c_{2} < 0$) at $T= 1\mu$K and $n_{0} = 10^{14}$cm$^{-3}$: q=0 (solid, black), $q = 2|c_{2}|n_{0}$ (blue, dashed), $q = 4|c_{2}|n_{0}$ (green, dotted), $q = 4.5|c_{2}|n_{0}$ (red, thin). (b) Gapless, linear dispersion for small $k$ at $q = 4|c_{2}|n_{0}$. The solid black curve is the small $k$ expansion (\ref{eq:7}). (c) $q = 2|c_{2}|n_{0}$ dispersion in detail. The horizontal black line is $\omega^{2} = q(q - 4|c_{2}|n_{0})$, which is the gap predicted by the small $k$ expansion (\ref{eq:7}). (d) $q=0$ dispersion in detail. Horizontal dashed line at $\omega^{2} = -4c^{2}_{2}n^{2}_{0}$ indicates the most unstable mode. Vertical dotted line is at $k\Lambda_{T} = \frac{\sqrt{2 |c_{2}|c_{0}}n_{0}}{k_{B} T}$.}
\end{figure}

Linearizing about this state, and dropping the collision term, the Boltzmann equation can be written in Fourier space:
\begin{eqnarray}\label{eq:4}
(-\omega + \frac{\vec{k}\cdotp\vec{p}}{m})\tilde{\delta\textbf{F}} = \Big[\textbf{V}_{0}, \tilde{\delta\textbf{F}}\Big] + \Big[\tilde{\delta\textbf{V}}, \textbf{F}_{0}\Big] + \\\nonumber
\frac{1}{2}\Big\{\vec{k}~\tilde{\delta\textbf{V}},\vec{\nabla}_{p}\textbf{F}_{0}\Big\} 
\end{eqnarray}
where $\textbf{F}_{0}$ is the initial distribution and $\textbf{V}_{0}$ is the initial interaction potential. The quantities $\tilde{\delta\textbf{F}}$ and  $\tilde{\delta\textbf{V}}$ are Fourier transforms of the change in the distribution and interaction, defined in the usual way $\delta f(\vec{k}, p, \omega) = \int d \vec{r}dt e^{i(\vec{k}\cdotp\vec{r} - \omega t)} \delta f(p,r, t)$.  Inserting the form for the interaction potential (\ref{eq:3}), we obtain the equations of motion for the density and spin fluctuations. 

We find that four long wavelength ($|k|v_{T} \ll \Omega_{int}$), low frequency ($\omega \ll \Omega_{int}$), spin modes propagate in this system. These modes have a quadratic dispersion, characteristic of ferromagnetic systems, but have a $k=0$ offset. Two of the modes are transverse spin waves, and the other two are quadrupolar modes where the longitudinal magnetization fluctuates along with the transverse spin fluctuations ($\langle S_\mu S_\nu\rangle -\langle S_\nu S_\mu\rangle$).  

The spin modes are given by the equation $1 = -(c_{0} + c_{2})\int\frac{d\vec{p}}{(2\pi)^{3}}\frac{f_{0} + \frac{1}{2}\vec{k}\cdotp\vec{\nabla}_{p}f_{0}}{\omega +\Omega_{int} + q - \frac{\vec{k}\cdotp\vec{p}}{m}}$ \cite{landau}. The transverse spin modes have dispersion $\omega(k) = \frac{k^{2}v^{2}_{T}}{2 \Omega_{int}} - q$. The fact that this has a negative energy at $k=0$ signifies that in the presence of a quadratic Zeeman shift, the initial state is not the thermodynamic ground state. One can decrease the energy by rotating the spins to be transverse to the magnetic field. However the frequency stays real, meaning that the initial state is metastable, and transverse spin will not be spontaneously generated. 

Meanwhile, the quadrupolar modes have dispersion $\omega(k) = -4c_{2}n_{0} + \frac{k^{2}v^{2}_{T}}{2 \Omega_{int}}$.  In a gas with positive $c_{2}$, the frequency vanishes at a finite wave-vector, indicating a thermodynamic instability in the anti-ferromagnetic gas. Conversely for negative $c_{2}$ the spin-dependent contact interaction leads to a gap in the quadrupolar mode spectrum.

As pointed out by Leggett \cite{leggett}, one can also derive these equations by writing the macroscopic equations for the density and spin current: 
\begin{eqnarray}\label{eq:5}
\partial_{t}\textbf{n} + \frac{\vec{\nabla}_{\textbf{R}}\cdotp\vec{\textbf{j}}}{m} =  i\Big[\textbf{V},  \textbf{n} \Big] \\\nonumber
\partial_{t}\vec{\textbf{j}} + \frac{\vec{\nabla}_{\textbf{R}}\textbf{Q}}{m} = i\Big[\textbf{V}, \vec{\textbf{j}} \Big] - \frac{1}{2}\Big\{\vec{\nabla}_{\textbf{R}}\textbf {V},\textbf{n} \Big\} - \frac{\vec{\textbf{j}}}{\tau_{D}}
\end{eqnarray}
where the energy density $\textbf{Q} = \int d{\vec{p}}~(\vec{p}\cdotp\vec{p})~\textbf{F}$. In order to obtain a closed set of equations, we approximate $\textbf{Q} \approx \int d{\vec{p}}~\frac{1}{3}\langle\vec{p}\cdotp\vec{p}\rangle~\textbf{F} = \frac{p_{T}}{2}\textbf{n}$, where $p_{T} = mv_{T}$. In a collisionless gas, sufficiently close to equilibrium, these approaches are equivalent. 

An advantage of the Leggett approach is that it provides access to the dispersion at large $k$ more readily than the former approach. Solving (\ref{eq:5}), we get the standard relations $\omega(k) = q - \frac{k^{2}v^{2}_{T}\tau_{D}/2}{(1 + (\Omega_{int}\tau_{D})^{2})} (\Omega_{int}\tau_{D} -  i)$, and $\omega(k) = -4c_{2}n_{0} + \frac{k^{2}v^{2}_{T}\tau_{D}/2}{(1 + (\Omega_{int}\tau_{D})^{2})} (\Omega_{int}\tau_{D} -  i)$. Setting $\tau_{D}$ to infinity, we once again recover the relationships derived above. 

\subsection{Excitations about the $m_{F} = 0$ state} Next we consider the case of a gas in the $m_{F} = 0$ state. As before, low energy dynamics are driven by a combination of the quadratic Zeeman energy, and the spin-dependent contact interaction. As we now show, in the case where $c_{2}$ is negative, these energies compete, driving a {\textit{dynamic}} instability. 
Linearizing equation (\ref{eq:4}) yields the relation:
\begin{equation}\label{eq:6}
1 = -\frac{4c^{2}_{2}\chi_{1}\chi_{2}}{(1+ (c_{0}+c_{2})\chi_{2})(1-(c_{0}+c_{2})\chi_{1})}
\end{equation}
where the response functions are $\chi_{1} = \int \frac{d\vec{p}}{(2\pi)^{3}}\frac{(f_{0} + \frac{1}{2}\vec{k}\cdotp\vec{\nabla}_{p}f_{0})}{-\omega +\Omega_{d}+ \frac{\vec{k}\cdotp\vec{p}}{m}}$ and $\chi_{2} = \int \frac{d\vec{p}}{(2\pi)^{3}}\frac{(f_{0} - \frac{1}{2}\vec{k}\cdotp\vec{\nabla}_{p}f_{0})}{-\omega -\Omega_{d}+ \frac{\vec{k}\cdotp\vec{p}}{m}}$ and $\Omega_{d}  = (c_{0}-c_{2})n_{0}-q$. For $|k|v_{T} \ll c_{0}n_{0}$ and $\omega \ll c_{0}n_{0}$.  The resulting dispersion is:
\begin{equation}\label{eq:7}
\omega^{2}(k) = q(q + 4c_{2}n_{0}) + (q + 2c_{2}n_{0})\frac{k^{2}v^{2}_{T}}{c_{0}n_{0}}
\end{equation}

\begin{figure}[hbtp]
\begin{picture}(50, 100)
\put(-60, 3){\includegraphics[scale=0.55]{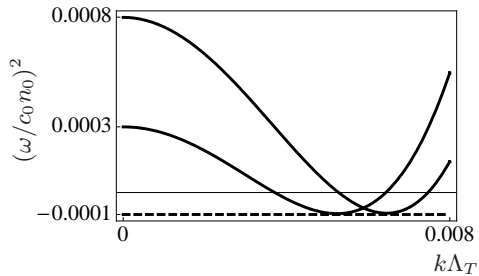}}
\put(-70, 35){\begin{sideways}\textbf{$(\omega/c_{0}n_{0})^{2}$}\end{sideways}}
\put(90, -6){\textbf{$k \Lambda_{T}$}}
\end{picture}
\caption{\label{fig:-2} Microwave-field induced instability in a polar gas for $q<0$: The modes of a polar gas with ferromagnetic interactions ($c_{2} < 0 $) are plotted for $q = 4c_{2}n$ (top) and $q = 2c_{2}n$ (bottom) (hence $q < 0$). The gaps are given by $q(q + 2c_{2}n_{0})$ in each case. Note that for negative $q$, a finite wave-vector instability sets in. Increasing $|q|$ pushes the instability to larger wave-vectors.  The dashed line is the most unstable frequency for $q=0$ (see Fig.~\ref{fig:-1}(d)) given by $\omega^{2} = - (2c_{2}n_{0})^{2}$. The timescale for the onset of the instability is independent of $q$. We set $T = 1\mu$K and $n_{0} = 10^{14}$cm$^{-3}$.}
\end{figure}  

The spin and quadrupole modes in the $m_{F} = 0$ state are degenerate.  For $c_{2} > 0$, or for large enough $q$, one obtains a gapped quadratic mode. For $c_{2} <0$ however, the dispersion goes from quadratic to linear  at $q=4|c_{2}|n_{0}$. Upon lowering $q$ further, the $m_{F} = 0$ state undergoes a {\textit{dynamic}} instability where the excitation frequency is complex. At $q=2|c_{2}|n_{0}$, the leading order term in the small $k$ expansion vanishes. 

The structure of the modes should be contrasted with those in the $m_{F} = 1$ state. Even though either case may not be a thermodynamically stable starting point for anti-/ferromagnetic interactions (respectively), only in the antiferromagnetic case do small fluctuations grow exponentially. 

Taking the $\tau_{D} \rightarrow \infty$ limit and solving (\ref{eq:5}), the dispersion relation is determined by the equation:
\begin{widetext}
\begin{eqnarray}\label{eq:8}
\Big((\omega - \Omega_{s})(\omega + \Omega_{d}) - \frac{v^{2}_{T}k^{2}}{2}\Big)\Big((\omega + \Omega_{s})(\omega  - \Omega_{d}) - \frac{v^{2}_{T}k^{2}}{2}\Big) = -4(c_{2}n_{0})^{2}(\omega - \Omega_{d})(\omega + \Omega_{d})\end{eqnarray}\end{widetext}
where $\Omega_{s} = q + 2c_{2}n_{0}$. Expanding (\ref{eq:8}) to lowest order in $k$ returns Eq.(\ref{eq:7}). 

The roots of (\ref{eq:8}) for negative $c_{2}$ are plotted in Fig.~\ref{fig:-1}(a) for various values of the quadratic Zeeman energy $q$.  The temperature is set to $1\mu$K, density to $10^{14}$ cm$^{-3}$, and we use the scattering lengths for $^{87}$Rb, $a_{0}(a_{2}) = 5.39(5.31)$nm. For the homogenous gas in consideration here, this is well above the temperature for Bose-Einstein condensation. As shown in Fig.~\ref{fig:-1}(b), at $q=0$, $\omega^{2}(k)$ is negative for small $k$, indicative of an instability. 

The timescale for the onset of this instability ($t_{ins}$) is determined by the most unstable mode, which from (\ref{eq:8}) occurs at $\omega^{2} = - (2 |c_{2}|n_{0})^{2}$, giving $t_{ins} \sim \frac{1}{2|c_{2}|n_{0}}$.The corresponding wave-vector is $k_{ins}\Lambda_{T} =  \frac{\sqrt{2 |c_{2}|c_{0}}n_{0}}{k_{B} T}$.  Solving for where the mode frequency vanishes, we find that all wave-vectors $k$ smaller than a critical wave-vector $k_{c} =  \sqrt{2}k_{ins}$ lead to unstable modes. For achievable experimental densities ($\sim 10^{14}$cm$^{-3}$) and temperatures ($\sim 1\mu$K), this wave-length of the instability $\lambda_{c} \sim 100\Lambda_{T} = 10\mu$m, which is comparable to typical cloud sizes. The time for the onset of this instability $\sim 0.25$s. Although this timescale is easily achievable in experiment, small collisional timescales ($\sim 50$ms) at these temperatures may make the instability undetectable in experiment.

\textit{Microwave induced instability in a ferromagnetic polar gas}:$-$
\footnote{This work was motivated by discussions with Mukund Vengalattore.}
Gerbier \textit{et al.} \cite{gerbier} have shown that a weak, microwave driving field, off resonant with the $F=1 \rightarrow F = 2$ transition, may be used to tune the energy levels of the $F=1$ hyperfine sublevels. In this sense, it plays the same role as the quadratic Zeeman shift. By changing the detuning, one can change the effective $q$  from positive ($E_{m_{F} = 0} < E_{m_{F} = \pm1}$) to negative ($E_{m_{F} = 0} > E_{m_{F} = \pm1}$), where $E$ denotes the energy of the hyperfine sublevels. 

It is therefore of experimental and theoretical interest to consider what happens to the $m_{F} = 1, 0$ states at negative $q$. For the $m_{F} =1$ state, the spin mode becomes gapped by the quadratic Zeeman energy, and the thermodynamic instability disappears. This is to be expected as a negative $q$ favors a ground state with $\langle \textbf{S}\cdotp\textbf{S} \rangle = 1$. 

The polar state however is more interesting. From (\ref{eq:7}), it is clear that for ferromagnetic interactions ($c_{2} < 0$), the first term on the right hand side is always positive, and spectrum becomes gapped. The second term however is negative, signaling a finite wave-vector instability. Analyzing (\ref{eq:8}) (see Fig.\ref{fig:-2}), we find that increasing $|q|$ pushes the instability to larger wave-vectors, but frequency of the most unstable mode remains largely unchanged. The experimental consequence is that $q<0$, one should observe a reduction in the size of ferromagnetic domains, while leaving the timescale for the onset of the instability unaffected.

\begin{figure*}[hbtp]
\begin{picture}(150, 90)(10, 10)
\put(-165, 100){(a)}
\put(-35, 100){(b)}
\put(95, 100){(c)}
\put(220, 100){(d)}
\put(-160, 15){\includegraphics[scale=0.375]{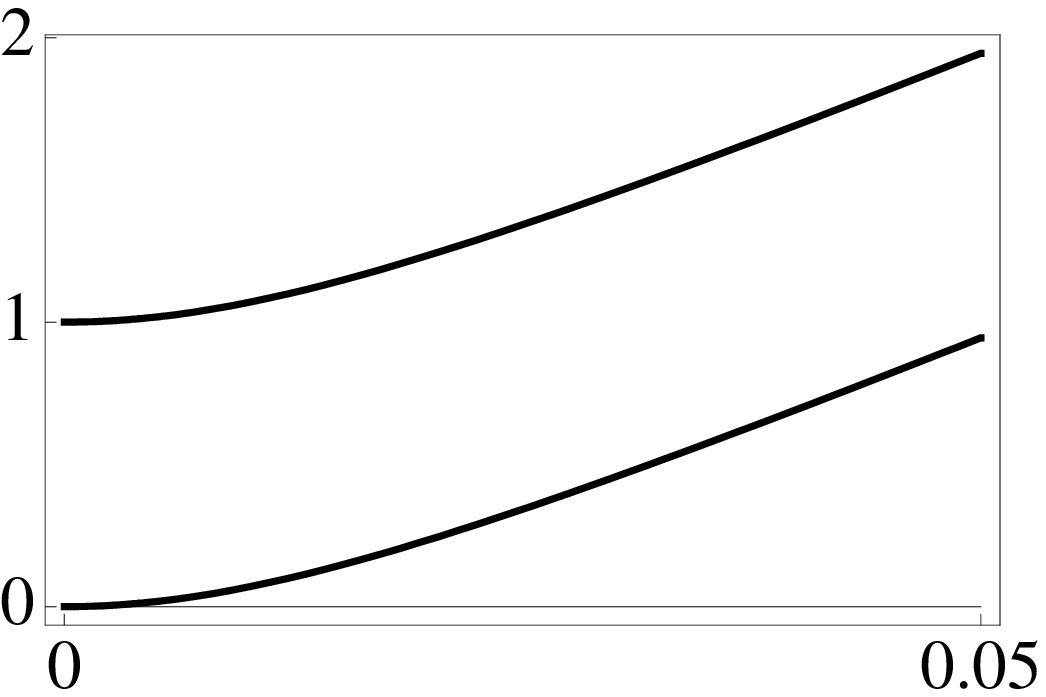}}
\put(-35, 15){\includegraphics[scale=0.39]{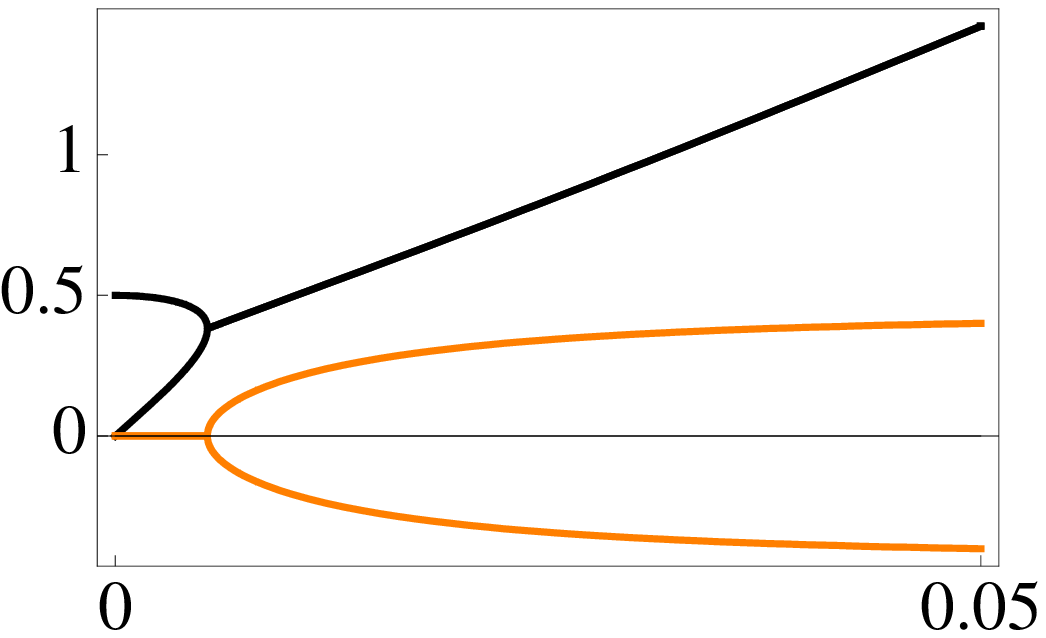}}
\put(95, 15){\includegraphics[scale=0.39]{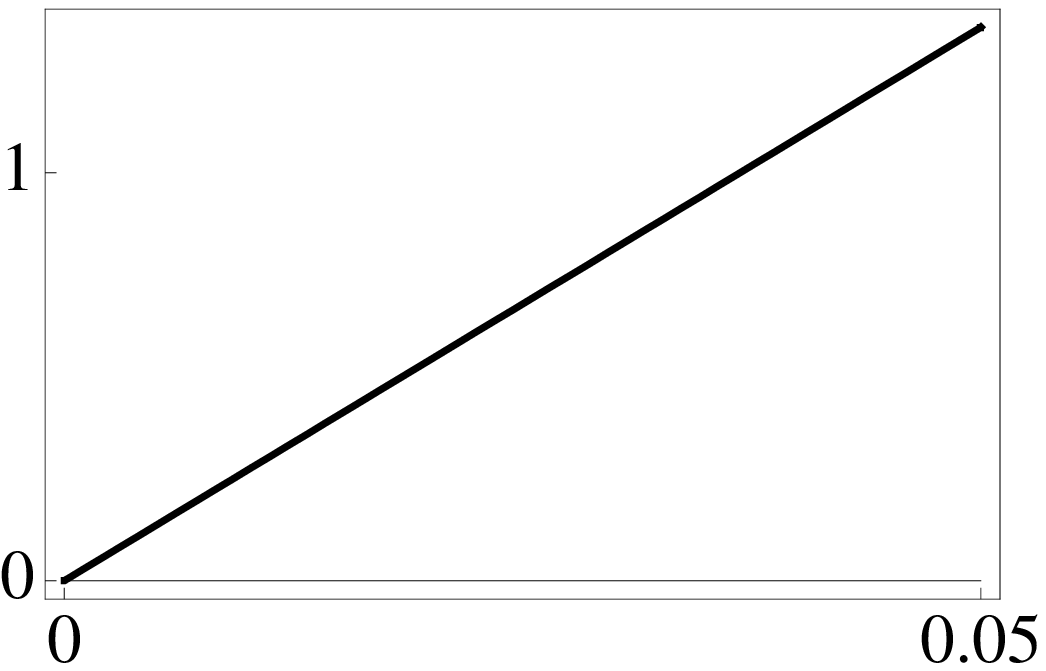}}
\put(220, 15){\includegraphics[scale=0.39]{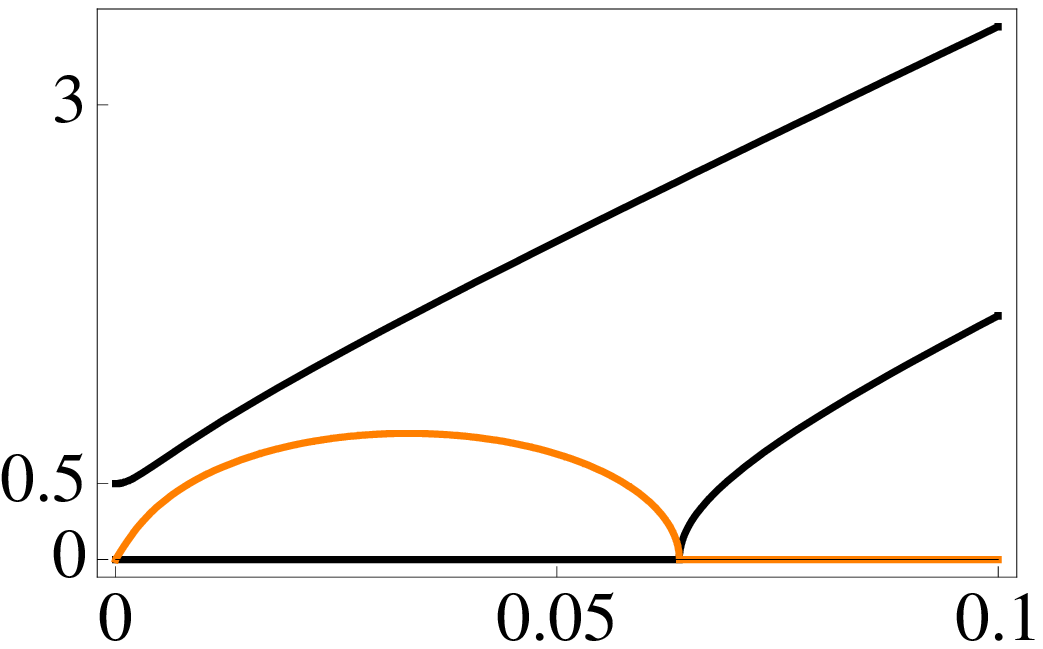}}
\put(-45, 42){\begin{sideways}\textbf{$\omega/c_{0}n_{0}$}\end{sideways}}
\put(84, 42){\begin{sideways}\textbf{$\omega/c_{0}n_{0}$}\end{sideways}}
\put(213, 42){\begin{sideways}\textbf{$\omega/c_{0}n_{0}$}\end{sideways}}
\put(-170, 42){\begin{sideways}\textbf{$\omega/c_{0}n_{0}$}\end{sideways}}
\put(200, 7){\textbf{$k \Lambda_{T}$}}
\put(70, 7){\textbf{$k \Lambda_{T}$}}
\put(-60, 7){\textbf{$k \Lambda_{T}$}}
\put(325, 7){\textbf{$k \Lambda_{T}$}}
\end{picture}
\caption{\label{fig:-3}(Color Online) Dispersion relations ($\omega(k)$) for $c_{2} > 0$ and $\sim c_{0}$. (a) Gapped (top) and ungapped (bottom) quadratic dispersion relation for $c_{2} = 0$.  (b) The two solid curves in (a) are now shown for $c_{2} = 0.5 c_{0}$ At some wave-vector the two modes become one single mode. The real (black) and imaginary (orange) components are indicated separately. The gap is equal to $\Omega_{d} = c_{0} - c_{2}$. (c) Linear dispersion $\omega(k) =  \frac{1}{\sqrt{2}}v_{T}k$, when $c_{2}/c_{0} = 1$. (d) Two modes for $c_{2} = 1.5c_{0}$: Gapped quadratic real mode; an unstable mode: real (black) and imaginary (orange) parts shown.  Notice that the instability only occurs if $0 < k < k_{max}(c_{2}/c_{0})$. The parameters for these plots are $T = 1\mu$K, $n \sim 10^{14}$cm$^{-3}$. For definiteness, we picked the atoms to have the mass of $^{87}$Rb and the corresponding $c_{0}$ value. We caution however that this choice is artificial, as $|c_{2}| \ll c_{0}$ in $^{87}$Rb. All these plots were made in zero field ($q=0$).}
\end{figure*}

\textit{Large anti-ferromagnetic interactions}: The modes of the polar gas, when $c_{2} > 0$ can be obtained from (\ref{eq:8}). For small $c_{2}/c_{0}$, the polar state is stable toward spin and nematic fluctuations. However, when $c_{2}$ and $c_{0}$ are commensurate, an instability can set in. This latter limit, which has been largely unexplored, is primarily of academic interest as all current spinor gas experiments have $|c_{2}| \ll c_{0}$. For simplicity we restrict our analysis to the case of no magnetic field ($q=0$).

As shown in Fig.~\ref{fig:-3}(a), for $c_{2} = 0$ we find a gapped and ungapped quadratic mode with dispersions $\omega(k) = \sqrt{\frac{1}{2}({\Omega^{2}_{d} \pm \sqrt{ \Omega^{4}_{d} - (v_{T}k)^{4}}})}$, where $\Omega_{d}$ reduces to $c_{0}n_{0}$ in this limit. As $c_{2}$ is increased, for some $k_{c}$, the gapped an ungapped modes meet, and for $k > k_{c}$, imaginary frequencies appear (Fig.~\ref{fig:-3}(b)). While the exact expression for $k_{c}$ is complicated, and depends on $T$ and density, we estimate the appearance of imaginary wave-vectors when $\Omega_{s} \approx \Omega_{d}$ i.e. $c_{2} \approx c_{0}/3$. The case of $c_{2} = c_{0}$ is special, and there is a single linear mode with frequency $\omega(k) =  \frac{1}{\sqrt{2}}v_{T}k$ (Fig.~\ref{fig:-3}(c)). However as $c_{2} > c_{0}$ (Fig.~\ref{fig:-3}(d)), imaginary frequencies appear once again for a bounded set of $k$ values: $0 < k < k_{max}(c_{2}/c_{0})$.

\subsection{Summary of Results}
In Fig. \ref{fig:-4} we summarize the results of our stability analysis in terms of the interaction parameters $c_{0}$ and $c_{2}$ for zero $q$. The following general features are clear:

\begin{figure}[hbtp]
\begin{picture}(50, 100)(10, 10)
\put(-80, 13){\includegraphics[scale=0.75]{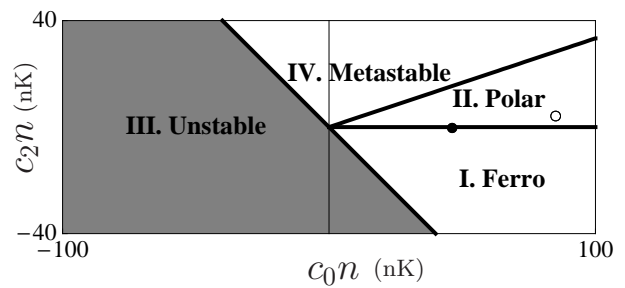}}
\put(30,7){\Large{$c_{0}n$}}
\put(55, 8){\small{(nK)}}
\put(-80, 47){\Large{\begin{sideways}$c_{2}n$\end{sideways}}}
\put(-81, 70){\small{\begin{sideways}(nK)\end{sideways}}}
\end{picture}
\caption{\label{fig:-4} Stability phase portrait in terms of the interaction strengths $c_{0}n$ and $c_{2}n$ at zero Zeeman field. The interaction strengths are measured in nK to facilitate comparison with typical experimental values. The four distinct regions are demarcated by solid black lines. I: For $c_{2} < 0$, but $c_{2} > -c_{0}$, the ferromagnetic state is the thermodynamic ground state. II: For $c_{2} > 0$, $c_{2} \le c_{0}/3$, the polar state is the ground state. III: Here $a_{2} (= c_{0} +c_{2}) < 0$, and the gas is dynamically unstable towards collapse. IV: When $c_{2} \ge c_{0}/3$, the polar state becomes metastable. In the presence of finite populations of $\pm1$, phase separation is expected to occur. In this region, the ferromagnetic state is thermodynamically unstable. Finally, the open and filled circles are typical interaction scales in $^{23}$Na and $^{87}$Rb for $n \sim 10^{14}$cm$^{-3}$.}
\end{figure}

\begin{enumerate}

\item For $c_{0} > 0$ and $c_{2}< 0$, the ground state is a ferromagnet. 

\item For $c_{0} > 0$ and $c_{2} > 0$, but $c_{2} \le c_{0}/3$, the ground state is polar. 

\item In region III the scattering length in the spin-$2$ channel ($a_{2}$) is negative, and the gas becomes mechanically  unstable.

\item In region IV, the polar state is metastable, in the presence of finite populations of $\pm1$ components, the system will phase separate \cite{mueller2}. The ferromagnet is thermodynamically unstable here.

\item The regime of current day experiments on spin-$1$ gases is that of $|c_{2}| \ll c_{0}$, indicated by the open and filled circles in the figure. For $^{87}$Rb, at $n \sim 10^{14}$cm$^{-3}$, $c_{0}n \sim 50$nK and $|c_{2}|n \sim 0.25$nK (filled). For $^{23}$Na, at the same density $c_{0}n \sim 85$nK and $c_{2}n \sim 4.25$nK (open).

\end{enumerate}

\section{Numerical Investigation} 

In an experiment, the atoms are typically confined to a harmonic trap. Calculating the modes in a trap is much more difficult than in free space. Example calculations \cite{nikuni} do not always yield simple physical pictures. Moreover experiments are typically performed in a limit where linear response is not applicable \cite{du}. 

For these reasons, we perform a numerical study of the spin dynamics for a trapped gas.  Furthermore while the calculations in part I pertained to the collisionless regime, using a simple relaxation time approximation, we model collisions in the gas. 

We address two questions here: 
\begin{enumerate}
\item What happens if the initial state is not a stationary state with respect to the external magnetic field ?
\item Can the instability in the polar state be observed experimentally?
\end{enumerate}

To address question $1$, we consider a ferromagnet in the $x$ direction, in the presence of a field along $z$. We  find that of coherent population dynamics in a collisionless gas. By controlling the rate of decay of spin current, one may be able to experimentally observe coherent oscillations even in a classical spinor gas. 

With regards to $2$, taking into account realistic experimental parameters, the polar state instability should not be visible in $^{87}$Rb. The collision time is set by $c_{0}$, while the time for spin dynamics by $c_{2}$. Owing to the small ratio of $|c_{2}|/c_{0} \sim 0.005$, collisions lead to relaxation before any coherences can develop. 

\subsection{Numerical setup}

We consider parameters such that the all the motion takes place in one dimension producing an effective 1D Boltzmann equation. Assuming that the distribution function in the transverse directions is frozen into a Boltzmann form, we can integrate them out  \cite{natu}, yielding renormalized scattering lengths ($a_{0}/a_{2}$). In what follows, we normalize all lengths by the oscillator length $\sqrt{1/(m\omega_{z})}$ and momenta by $\sqrt{m k_{B} T}$. Using a phase-space conserving split-step approach \cite{teuk}, we numerically integrate (\ref{eq:2}). We have verified that the tempero-spatial grid is fine enough that  our results are independent of step-size. 

We use a radial trapping frequency $\omega_{r} = 2\pi \times 250$Hz, axial trapping frequency of $\omega_{z} = 2\pi \times 50$Hz, roughly $5 \times 10^{5}$ atoms, and set the temperature to $1\mu$K. The initial distribution function in position and momentum is given by the equilibrium distribution $e^{-\beta(\frac{p^{2}}{2m} + \frac{1}{2}m\omega^{2}_{z}z^{2})}$. 

\subsection{Ferromagnetic gas in the x-direction}

\begin{figure}[hbtp]
\begin{picture}(100, 80)(10, 10)
\put(-48, 87){(a)}
\put(60, 87){(b)}
\put(60, 15){\includegraphics[scale=0.37]{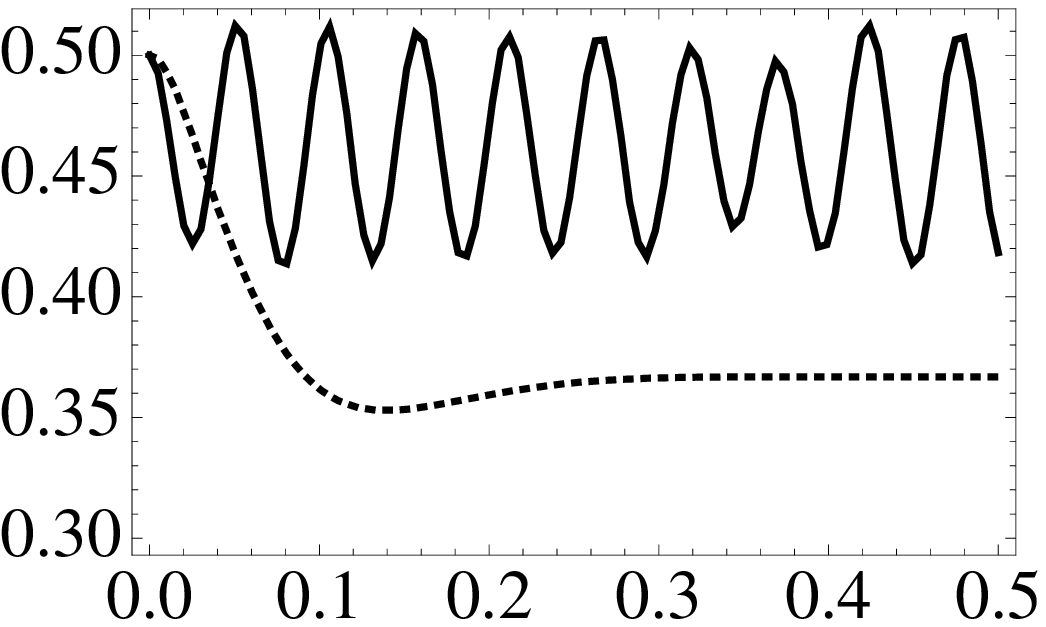}}
\put(-57, 15){\includegraphics[scale=0.35]{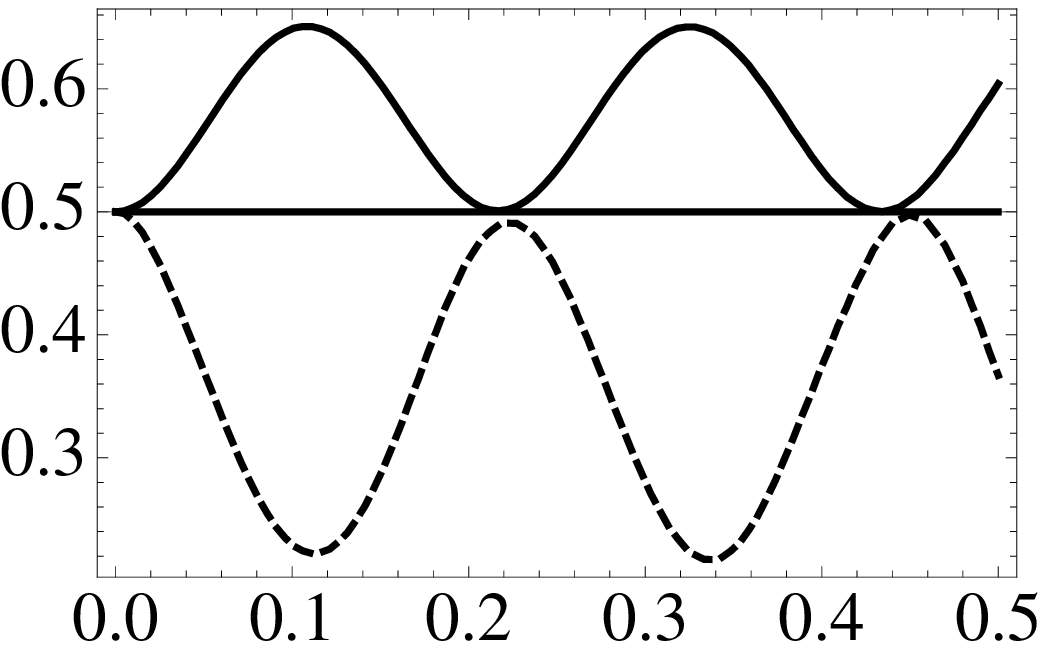}}
\put(50, 40){\begin{sideways}\textbf{$n_{00}/n_{0}$}\end{sideways}}
\put(-68, 40){\begin{sideways}\textbf{$n_{00}/n_{0}$}\end{sideways}}
\put(160, 08){\textbf{$t(s)$}}
\put(40, 08){\textbf{$t(s)$}}
\end{picture}
\caption{\label{fig:-5}(a) Evolution of central density of $m_{F} = 0$ component (normalized to initial density $n_{0} = n_{00} + n_{11} + n_{-1-1} (t = 0)$) as a function of time for negative $c_{2}$ ($^{87}$Rb parameters (thick line)) and positive $c_{2}$ ($^{23}$Na parameters (dashed curve)) for $q \sim 15$Hz. ($B \sim 0.1$G).  The line at $n_{00}/n_{0} = 0.5$ is a guide to the eye. (b) Thick curve: evolution of $m_{F} = 0$ population in $^{23}$Na for larger $q \sim 60$Hz ($B \sim 0.3$G). Dashed curve: suppression of oscillations due to finite relaxation rate ($\tau_{el} \sim 80$ms) in  $^{23}$Na within a relaxation time approximation. Parameters used in the simulations: $\omega_{r} = 2\pi \times 250$Hz, $\omega_{z} = 2\pi \times 50$Hz and $N =5 \times 10^{5}$.} 
\end{figure}

The spin density matrix for a ferromagnet pointing in the $x-$ direction is: $\textbf{F} = \psi^{*}\psi$, where $\psi = \{\frac{1}{2}, \frac{1}{\sqrt{2}}, \frac{1}{2}\}$. Numerically integrating (\ref{eq:5}), we find that for a magnetic field of $0.1$G ($q \sim 15$Hz), in the absence of collisional relaxation, the populations of the three hyperfine sublevels coherently oscillate about their equilibrium values, as shown in Fig.~\ref{fig:-5}(a). The amplitude of the oscillations depends on the magnitude of the spin dependent contact interaction. The evolution is not simply a precession, rather it involves an oscillation between ferromagnetism and nematicity.

For $c_{2} < 0 (> 0)$ the population of the $m_{F} = 0$ sublevel oscillates between $0.5$ and a larger (smaller) value. The frequency and amplitude of the oscillations also depend strongly on the magnitude of the quadratic Zeeman energy. Increasing $q$ increases the frequency of oscillations, but decreases the amplitude (Fig.~\ref{fig:-5}(b)). Similar results have been obtained by Chang \textit{et al.} \cite{chapman2} in condensed $^{87}$Rb, although they considered a different initial spin configuration. 

The coherences oscillate at a frequency set by the external perturbation, in this case the quadratic Zeeman energy. The resulting phase factors can be interpreted as rotations in spin space. The spin-dependent contact interaction couples spin and density, and oscillations in the densities of the three components are seen. Increasing $q$, causes the spin vector to rotate faster, leading to a rapid averaging to its initial value over the timescale of the spin-dependent contact interaction, reducing the size of the effect.

\textit{Collisions}:$-$ Experiments will also involve collisions between the atoms. A detailed derivation of collisional relaxation rates can be found in \cite{nikuni}. We are concerned here with the experimental feasibility, hence we use a simple relaxation time approximation. The relaxation rate is proportional $(\frac{a_{0}+ 2a_{2}}{3})^{2} \sim a^{2}_{0}$. The amplitude of the oscillations in the populations of the three sub-levels is directly proportional to the size of $c_{2}$. 

From the trap dimensions, particle number, and the scattering lengths for $^{87}$Rb ($|c_{2}|/c_{0} \sim 0.005$), we estimate $\tau_{el} \sim 20$ms. On this timescale, virtually no oscillations are seen. The story is different for $^{23}$Na where the smaller scattering length implies relaxation times almost 4 times longer than in Rb, and the ratio of $|c_{2}|/c_{0}$ is $10$ times larger. As shown in Fig.~\ref{fig:-5}(a,b) an experiment with sodium atoms should be able to detect population dynamics.

\subsection{Polar state instability}

\begin{figure}[hbtp]
\begin{picture}(100, 150)(10, 10)
\put(-55, 170){(a)}
\put(68, 170){(b)}
\put(-55, 78){(c)}
\put(68, 78){(d)}
\put(68, 5){\includegraphics[scale=0.36]{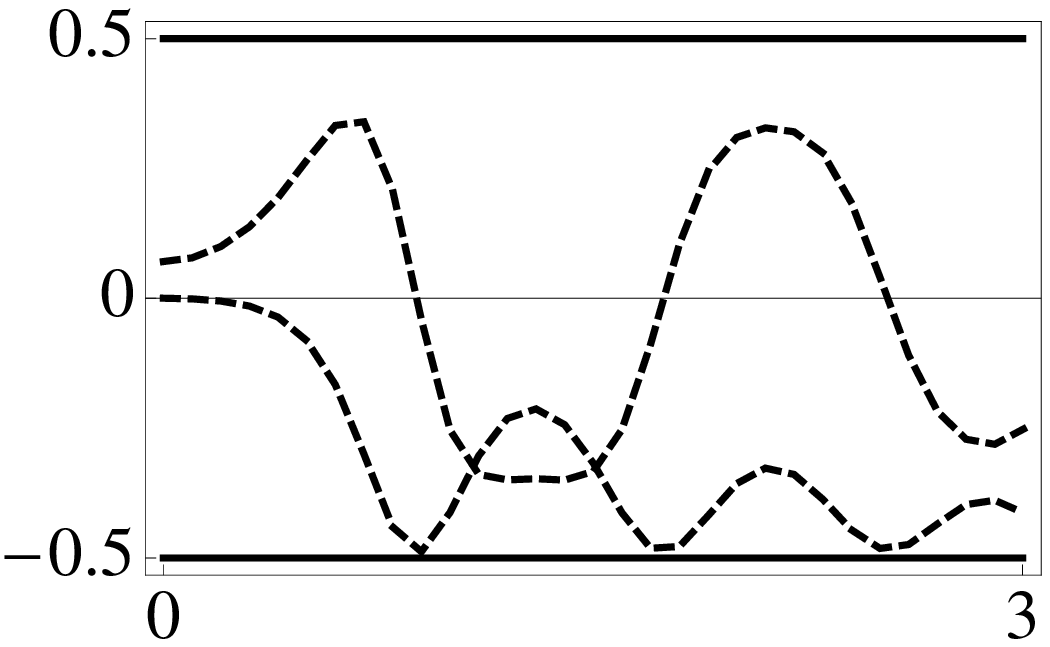}}
\put(-50, 5){\includegraphics[scale=0.35]{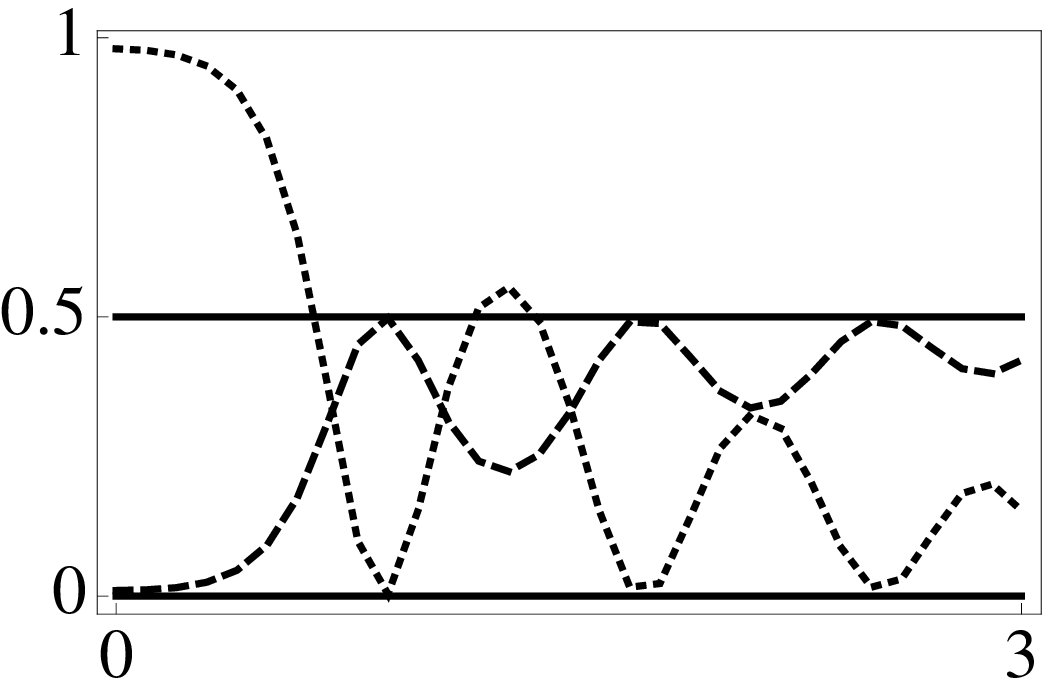}}
\put(68, 95){\includegraphics[scale=0.37]{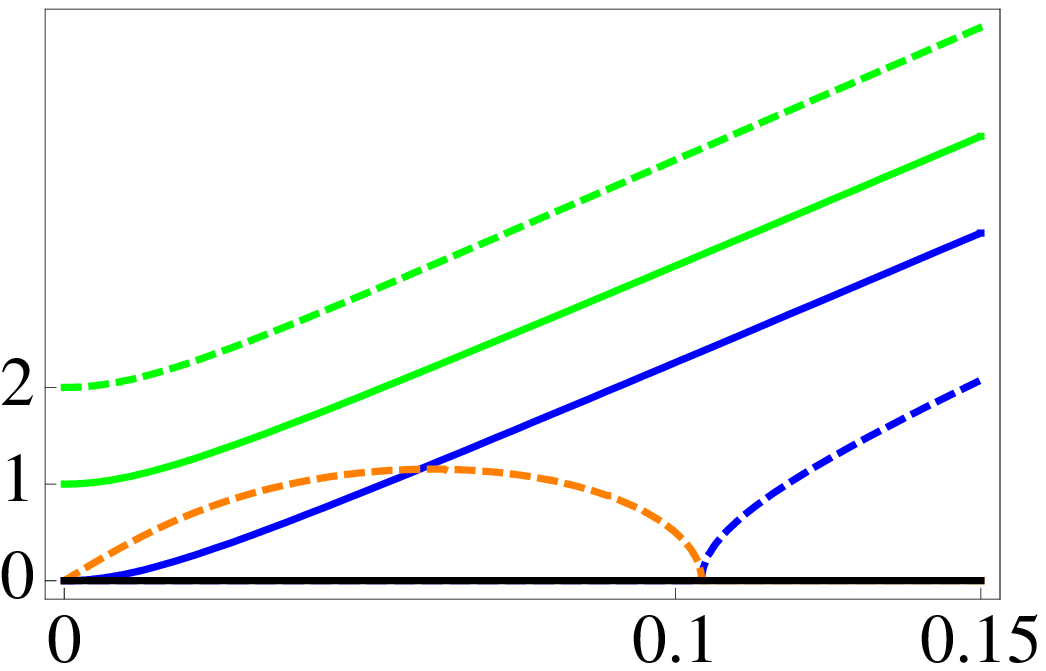}}
\put(-55, 95){\includegraphics[scale=0.37]{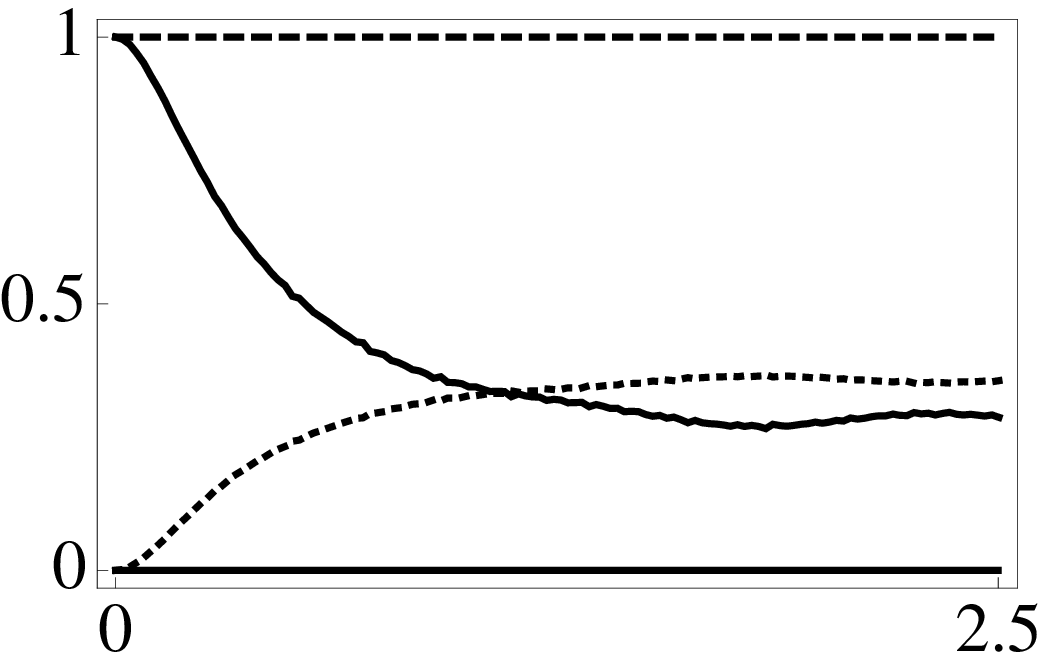}}
\put(62, 135){\begin{sideways}\textbf{$\omega/c_{0}n_{0}$}\end{sideways}}
\put(-65, 125){\begin{sideways}\textbf{$n_{jj}/n_{0}$}\end{sideways}}
\put(165, 87){\textbf{$k\Lambda_{T}$}}
\put(45, 87){\textbf{$t(s)$}}
\put(63, 34){\begin{sideways}\textbf{$n_{ij}/n_{0}$}\end{sideways}}
\put(-62, 34){\begin{sideways}\textbf{$n_{jj}/n_{0}$}\end{sideways}}
\put(165, 0){\textbf{$t(s)$}}
\put(45, 0){\textbf{$t(s)$}}
\end{picture}
\caption{\label{fig:-6}(a) Collisionless evolution of the $m_{F} = 0~(+1)$ densities (solid/dotted curves) for $c_{2} = -c_{0}$, for an initial state $\{\frac{\epsilon}{\sqrt{2}}, 1, \frac{\epsilon}{\sqrt{2}}\}$ (spin fluctuation), where $\epsilon = 0.01$. The dashed curve shows the $m_{F} = 0$ populations when $\psi_{initial} = \{\frac{\epsilon}{\sqrt{2}}, 1, -\frac{\epsilon}{\sqrt{2}}\}$ (nematic fluctuation). The densities are normalized to the total initial density $n_{0} = n_{00} + n_{11} + n_{-1-1}(t = 0)$.
(b) A gapped and ungapped (solid blue and green) exist for $c_{2} = 0$. When $c_{2} \rightarrow -c_{0}$, (dashed blue and green curves) the energy of the gapped mode increases (green dashed) while the other mode (blue dashed) becomes unstable. The dashed orange curve is the imaginary part of this mode. 
(c) Collisionless evolution of central density of $m_{F} = 0, +1$ components (dotted/dashed respectively) for negative $c_{2}$ ($c_{2}/c_{0} \sim 0.05$ and $c_{2} \sim 2$Hz) for $q \ll |c_{2}|n$ ($B \sim 1$mG). Conservation of $S_{z}$ implies that the $m_{F} = -1$ component evolves in the same way as $m_{F} = 1$. (d) Collisionless evolution of coherences - top curve is the $n_{0-1}$ component and bottom is the $n_{1-1}$ component normalized to the initial total density. The $n_{1-1}$ component tends to -0.5, coherence is maintained, and the final state is a polar state along $x$.}
\end{figure}

In Sec.II(B) we showed that the polar state of a gas with ferromagnetic interactions has a dynamically unstable mode at long wavelengths. Here we explicitly demonstrate this instability for a trapped gas with ferromagnetic interactions. However we also find that this instability is not observable in current experiments. Given the scattering lengths of $^{87}$Rb, our simulations indicate that the collision times are so short that the coherences necessary to drive this instability will never have time to develop.

Additionally we explore details of the instability. We verify that the polar state is always unstable for small $q$, but the final state is dependent on the magnitude of $|c_{2}|/c_{0}$. For small values of this ratio ($\sim 10^{-3}$), finite populations of $m_{f} =  \pm 1$ develop, but not all the atoms are transferred out of the $m_{F} = 0$ state. Increasing $c_{2}$ decreases the timescale for the onset of this instability, as well as increasing the fraction of the population transferred. The $m_{F} = 0$ state is stabilized at large Zeeman fields. Note that $S_{z}$ is always a conserved quantity.

We start with a polar gas with small coherences, which seed the instability. We consider seeds with two different symmetries: $\textbf{F} = \psi^{*}\psi$ where $\psi = \{\psi_{1}, \psi_{0}, \psi_{-1}\} = \{\frac{\epsilon}{\sqrt{2}}, 1, -\frac{\epsilon}{\sqrt{2}}\}$ (nematic)  or $\{\frac{\epsilon}{\sqrt{2}}, 1, \frac{\epsilon}{\sqrt{2}}\}$ (spin), where $\epsilon \ll 1$, and we only keep terms ${\cal{O}}(\epsilon)$ in the density matrix. The Wigner functions are assumed to have a Maxwellian form in phase space.  

First consider $q = 0$ dynamics. As the polar state is unstable to spin fluctuations, an initial spin fluctuation grows exponentially, while a nematic fluctuation merely changes the size of the nematic director, but does not make it unstable (Fig.~\ref{fig:-6}(a)). To enhance the size of the effect, we pick $c_{2} = -c_{0}$, but we observe an exponentially growing spin mode for typical values of $c_{2}$ as well.

Although the linear stability analysis predicts an instability, it does not determine the final state, which depends in a complicated way on the long time dynamics. In  Fig.~\ref{fig:-6}(a) we show the collisionless evolution of the $m_{F} = 0 (+1)$ populations (solid/dashed) starting in the state $\{\frac{\epsilon}{\sqrt{2}}, 1, \frac{\epsilon}{\sqrt{2}}\}$ with $\epsilon = 0.01$. The dashed curve shows that for an initial state with nematic order $\{\frac{\epsilon}{\sqrt{2}}, 1, -\frac{\epsilon}{\sqrt{2}}\}$, no dynamics is seen. 

In Fig.~\ref{fig:-6}(b) we plot the modes of the polar gas after solving (\ref{eq:8}). As before there are two modes at $c_{2} = 0$.  One mode has a gap of $(c_{0} - c_{2})n_{0}$ (green curves), which increases as $c_{2} \rightarrow -c_{0}$. The other ungapped mode (blue curves) develops an imaginary part (orange) which increases as $c_{2}$ is made more negative.

The quadratic Zeeman effect stabilizes the polar state. Given an initial spin fluctuation about the unmagnetized state, this spin precesses in the Zeeman field at a rate proportional to $q$. If the initial state only has a nematic perturbation, then a finite $q$ is needed to produce the spin fluctuations required to drive an instability. Upon increasing $q$, there are two effects. First there appear wave-vectors where the frequency is real (\ref{eq:8}), and oscillations are observed along with an exponential growth of spin populations. Second, the spin vector precesses more rapidly, and averages to zero on a timescale of $\sim |c_{2}|n$, ultimately stabilizing the polar state.

In Fig.~\ref{fig:-6}(c) we plot the evolution of the central density of the $m_{F} = 0, +1$ sub-levels in time for $q \sim 10$mHz, in a collisionless gas. We take $c_{2}n \sim 2$Hz, such that $|c_{2}|/c_{0} \sim 0.05$, instead of the $0.005$ in $^{87}$Rb. In Fig.~\ref{fig:-6}(d) the coherences are shown. The timescale for the onset of the instability, as well as the final state depends on the size of the off-diagonal seed, and the nature of the perturbation i.e. whether it is a spin or a nematic fluctuation or both, and the strength of the interactions. Here we have shown one final state, $\psi_{f}  = \{\frac{1}{\sqrt{2}}, 0,-\frac{1}{\sqrt{2}} \}$. Note that the final state is not simply an incoherent mixture, unlike in a collisional gas. The distribution functions for the $m_{F} = \pm1$ components are identical, and no local magnetization is observed. Local structures may be observed in an experiment where the trapping potentials are made spin dependent. Performing the simulation with a relaxationn rate approximation, taking into acccount collision times ($\tau \sim 50$ms), and scattering lengths for $^{87}$Rb, we do not observe any dynamics in the polar state.

\begin{figure}[hbtp]
\begin{picture}(50, 100)(10, 10)
\put(-50, 15){\includegraphics[scale=0.5]{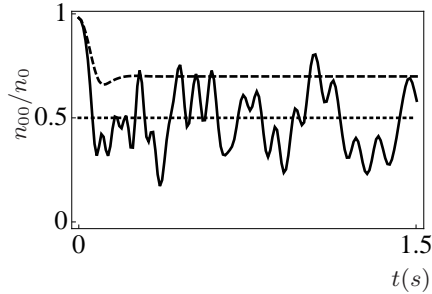}}
\put(85, 5){\textbf{$t(s)$}}
\put(-60, 60){\begin{sideways}\textbf{$n_{00}/n_{0}$}\end{sideways}}
\end{picture}
\caption{\label{fig:-7} Evolution of central density of $m_{F} = 0$ component (normalized to initial density $n_{0} = n_{00} + n_{11} + n_{-1-1} (t = 0)$) as a function of time for $c_{2} =  0.5 c_{0}$ without (solid) and with the relaxation rate approximation (dashed). The initial state is $\textbf{F} = \psi^{*}\psi$, where $\psi = \{\frac{\epsilon}{\sqrt{2}}, 1, \frac{\epsilon}{\sqrt{2}}\}$ (spin fluctuation), where $\epsilon = 0.01$. Trap parameters: $5 \times 10^{5}$ atoms, $\omega_{r} = 2\pi \times 250$Hz, $\omega_{z} = 2\pi \times 50$Hz. The relaxation time $\tau \sim 50$ms. For definiteness we assumed $^{87}$Rb atoms, even though the scattering lengths in Rb, do not obey this relationship. Owing to the strong interactions, one should be able to observe this instability even for reasonable collision rates.}
\end{figure}

\textit{Large anti-ferromagnetic interactions}: Next we consider the case where the spin-dependent and spin-independent contact interactions are comparable in magnitude, and both positive. In Fig.~\ref{fig:-7} we plot the central densities of the $m_{F} = 0$ atoms as a function of time for $c_{2} = 0.5c_{0}$ without (solid curve)  and with (dashed) the relaxation time approximation, demonstrating the instability. Due to the large interaction energies, the dynamics is complicated.

As Endo and Nikuni \cite{nikuni} have shown, one must be careful while considering the relaxation time approximation in the $c_{0} \sim c_{2}$ limit where more complicated spin dependent collisions may become important. In order to fully model the physics, more than one ``relaxation-rate" may be required. 

\section{Summary and Outlook} 

We have addressed the role of the spin-dependent interaction in the collisionless dynamics of a thermal spinor gas. By calculating spin and quadrupolar modes, about the magnetized and unmagnetized states, we have shown that the normal state of a spinor has a rich array of spin excitations compared to its well-studied pseudo-spin $\frac{1}{2}$ counterpart.  We numerically calculated the dynamics of a non-stationary initial state, finding that the spin-dependent contact interaction drives population dynamics. Finally we provide an explicit demonstration of the instability of the polar state to spin fluctuations.  

We conclude that many interesting experiments may be done in the \textit{normal} state of a spinor gas. In order to observe some of the physics described here, it will be necessary to attain a limit where the spin dependent  and independent interactions are commensurate in magnitude. We hope that this work motivates experiments in that direction.

\textit{Acknowledgements}:$-$ S.N. would like to thank Stefan K. Baur and Kaden R.A Hazzard  for discussions. We thank Mukund Vengalattore for pointing out \cite{gerbier} to us, for several helpful discussions and comments on this manuscript. This work was supported by NSF Grant No.~PHY-0758104.

\end{document}